# Surface effect that causes a peak in band-edge photocurrent spectra: a quantitative model

*Yury Turkulets, Tamar Bick, and Ilan Shalish*

Ben Gurion University, Beer Sheva, 8410501, Israel

E-mail: shalish@bgu.ac.il

## Abstract

Band edge photocurrent spectra are typically observed in either of two shapes: a peak or a step. In this study, we show that the photocurrent band edge response of a GaN layer forms a peak, while the same response in GaN nanowires takes the form of a step, both are red-shifted to the actual band edge energy. Although this apparent inconsistency is not limited to GaN, the physics of this phenomenon has been unclear. To understand the physics behind these observations, we propose a model explaining the apparent discrepancy as resulting from a structure-dependent surface-effect. To test the model, we experiment with a GaAs layer showing that we can deliberately switch between a step and a peak. We demonstrate that using this quantitative model one may obtain the exact band edge transition energy, regardless of the red-shift variance, as well as the density of the surface state charges that cause the red shift. The model thus adds quantitative features to photocurrent spectroscopy.

Keywords: Spectral photoconductivity; Photocurrent spectroscopy; Exciton; Surface states; GaN; Nanowires; GaAs; Franz-Keldysh effect

**I. Introduction**

Spectroscopy of photocurrent or photoconductivity has been one of the most widely used tools in characterization of electro-optic materials and devices and in the study of light-semiconductor interaction. It is the basic method used to obtain detector spectral responsivity and quantum efficiency curves – a standard and actively used technique.[1] The theoretical foundation of the method is long established, and recently, it has seen a resurgent interest for nanostructure characterization.[2-10] Nonetheless, an important aspect of its interpretation appears to have been poorly understood.

Excluding relatively minor photoresponse of deep levels, the main response of a semiconductor starts at about the bandgap energy (often dubbed *intrinsic photoconductivity*).[11] The exact onset of the intrinsic photoresponse is of great interest to material scientists and engineers, because it bears on the ability to utilize the material for purposes such as light detection and solar energy conversion.[1] Yet, our experience shows that the onset energy may sometimes vary even among different spots on the same wafer. Moreover, the shape of this onset can vary, as we show here, between a step and a peak. As a result of these variations, this onset has often been regarded not reliable for measuring the bandgap or exciton energy. In this paper, we propose a model to explain the physics underlying the energy position and the shape of the band edge photocurrent spectrum. We also propose a method to extract the *exact optical transition energy* (band edge energy) as well as the *surface state density* that we find indirectly responsible for the variability of the onset.

**II. Model**

Photocurrent in semiconductors reflects an increase of the conductivity induced by photon absorption. Its spectra typically show a transition at the *absorption edge* that is about the bandgap energy. The spectral onset of the absorption edge generally precedes the band edge mainly due to an electric-field-assisted absorption.[3] This is because the built-in electric field, present at surface depletion regions, adds to the photon energy, assisting photons with energy smaller than the bandgap to excite electrons across the gap (schematically depicted in Fig. 1). Hence, the stronger the built-in field, the lower is the onset energy. The surface built-in field is caused by surface charges, trapped at surface states, and, therefore, depends on their concentration.[12] As a consequence of the typical variance in surface state density, the spectral data of absorption-related bandgap transitions has often been considered inaccurate for measuring the bandgap energy.[13]

As illustrated in Fig. 1, the lower the photon energy, the higher the distance needed to be tunneled in the forbidden gap, and accordingly, the lower is the probability of tunneling. As a result, a band-to-band absorption transition will always commence below the bandgap energy and increase gradually until the gap energy is reached. The tunnel-barrier may be approximated to be triangular. Using the WKB approximation, the probability of tunneling is

$$T(h\nu) = exp\left(-\left(\frac{E_g - h\nu}{\Delta}\right)^{3/2}\right) \quad (1)$$

where $E_g$– bandgap energy, $h\nu$– photon energy, and

$$\Delta = \left(\frac{3}{4}\frac{qE\hbar}{\sqrt{2m}}\right)^{2/3} \quad (2)$$

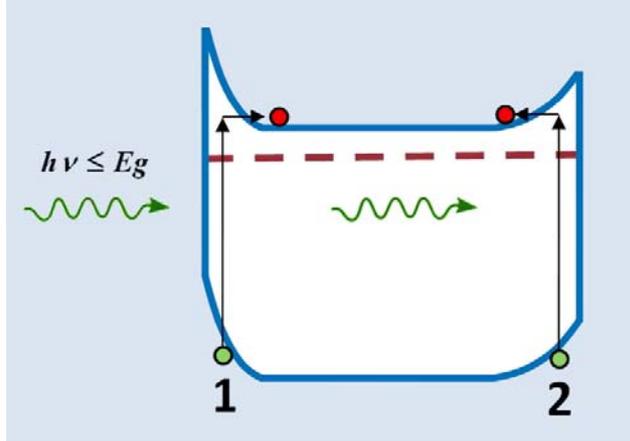

**Fig. 1** Band diagram of surface depleted n-type semiconductor layer irradiated with photons of energy slightly smaller than the bandgap. Due to the electric field near each surface, the bands are bent, which makes it possible for electrons to get from the valence band to the conduction band with less than the bandgap energy, by tunneling through the forbidden gap as shown. As the band edge energy is approached, the photon flux arriving at surface #2 will gradually decrease due to increasing absorption in the bulk of the layer.

$E$–surface electric field, $q$– electron charge, $m$– reduced effective mass, and $\hbar$– reduced Plank constant. This model was introduced independently by Keldysh and by Frantz to describe absorption of light below the bandgap.[14] This model is the basis for the electromodulation technique in photoreflectance spectroscopy.[15] It has been used by Franssen *et al*. to explain spectral behavior of photocurrent in InGaN quantum wells and by Cavallini *et al*. to assess bandgap energy in GaN nanowires.[16,3]

At sub-bandgap photon energies, a semiconductor is essentially transparent. Such photons will pass through the layer reaching the back side of the layer, and some will come out the back surface. As both top and bottom surfaces are typically depleted (Fig. 1), field-assisted band-to-band transitions can take place at the two surfaces (processes 1 and 2 in Fig. 1) already at photon energies lower than the gap. The resulting below-gap-photocurrent will therefore be a superposition of the effects of absorption on the conductivity at the two surfaces.

As the photon energy gradually approaches the band-edge, the back side of a layer will receive a gradually *decreasing* photon flux due to gradually increasing absorption of photons on the way, in the bulk of the layer. As a result, the photocurrent contributed by the *back surface* will gradually diminish and, eventually, be eliminated altogether. This decrease always starts before the actual bandgap energy is reached. As a result, the photocurrent will show a peak feature peaking *below* the actual band edge or exciton energy. Similar response of buried or back surfaces is well known in other absorption based spectroscopies, e.g., surface photovoltage spectroscopy.[17] We will now show a few experimental examples.

**III. Materials and Methods**

Si doped n-type GaN layers grown on sapphire were obtained from TDI Inc. Ti(5nm)\Au(100nm) contacts were e-beam evaporated and were unintentionally heated during the e-beam deposition. Their ohmic character was verified using current-voltage characteristics. The growth and preparation of GaN single nanowire (NW) devices has been described elsewhere.[18] During spectral acquisition a voltage of 0.1 V was applied between the contacts. Illumination was carried out using Xe arc lamp (for GaN) or halogen lamp (for GaAs) monochromatized by a double Newport MS257 spectrometer followed by long-pass filters to eliminate high order diffractions. The spectrometer was operated in a closed control loop to maintain a constant photon flux throughout the spectral range of the measurement. The wavelength was stepped from long to short wavelengths at equal energy steps. Each data point is an average of 100 consequtive measurements of the same point.

**IV. Results and Discussion**

Figures 2a and 2b show near band-edge photocurrent spectra of a single GaN nanowire and of a GaN film on sapphire, respectively. The nanowire shows a step-shaped response. The GaN layer shows an increase followed by an almost total decrease, with the peak centered at ~3.37 eV, about 50 meV below the GaN band-edge. The GaN film was grown on sapphire, thus the back surface, in this case, is the interface with the sappire substrate – a nucleation layer that is typically defective due to the lattice mismatch of the two crystals. The density of defects at the bottom surface is likely greater than that of the front surface, and the electrical conductance at the bottom is typically higher, resulting in a greater contribution to the photocurrent. This contribution is eventually eliminated by bulk absorption before the band edge energy is reached, and it seems that for this reason, the photocurrent almost totally subsides.

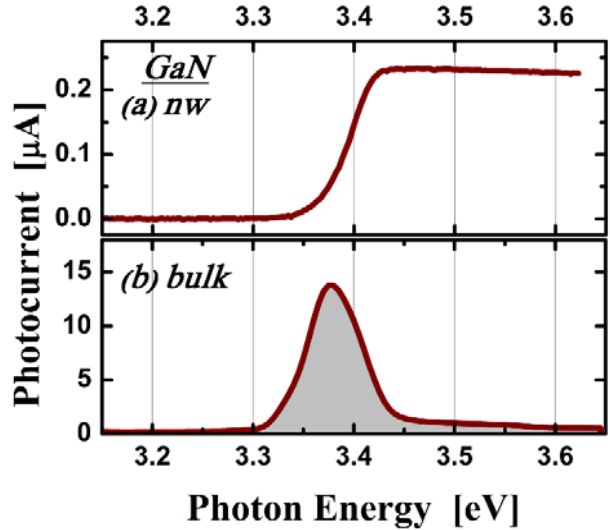

**Fig. 2** Photocurrent spectra obtained from (a) GaN nanowire showing a step, (b) GaN wafer showing a peak.

In the case of the nanowire, one may think of good reasons for a peak shape (e.g., density of states features in low dimensional structures).[19] Therefore, if there were a peak in the nanowire response, it would be difficult to identify the cause for this peak. The fact that it nonetheless follows a step is therefore in line with a simple argument that there is practically nothing that could cause a peak. Specifically, there is no bulk absorption between the nanowire surfaces, but this is expected as, practically, a nanowire has no bulk.

If one could eliminate the band bending at the back surface of a layer, one should be able to see a step in a layer as well. To test this hypothesis, we used a GaAs(100) wafer with a semi-insulating epilayer. As the epilayer is practically intrinsic, there is little to no built-in field at the epilayer surface. The main photocurrent is therefore due to absorption at the other surface, i.e., we have practically only a single surface contributing to the photocurrent. Two spectra were acquired: One, where illumination was from the epilayer side, and another from the opposite side. In the former, the responding surface was at the bottom, and the light had to traverse the bulk of the sample to reach it. In the latter, the absorbing surface was on top.

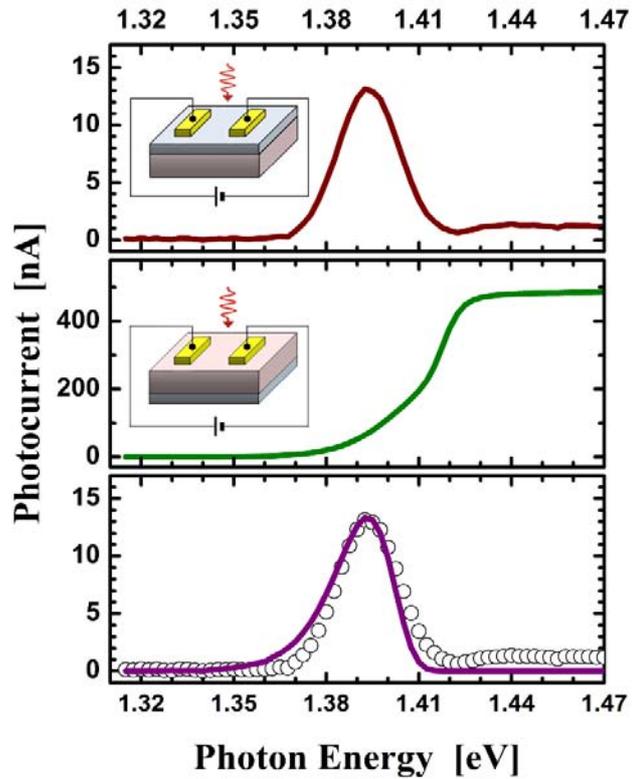

Figure 3a shows the photocurrent spectrum when the responding surface is at the bottom. As expected, the photocurrent shows a peak below the band-edge (at *1.37eV*, slightly below *Eg=1.42 eV*). Figure 3b shows the photocurrent spectrum when the responding surface is on top. As expected, the photocurrent rises, reaches a maximum and essentially remains constant thereafter. To further test the model, we attempted to calculate the first spectrum (the peak) from the second spectrum data (the step) by multiplying the spectrum with the factor *(1-T)·exp(-a(hv)·t)*, where *T*– transmission through the wafer, and *t*– wafer thickness (400 μm), and *a(hv)*–absorption coefficient as a function of photon energy taken from Casey *et al.*[20] Figure 3c shows the calculated spectrum (solid line) on top of the measured spectrum of Fig. 3a (circles). The small differences may be due to the fact that the absorption coefficient was not measured on our specific wafer, and also because we approximated the transmission to be constant, while this is actually another function of the photon energy. Nonetheless, the fit is still good enough to convince that the drop in photocurrent that follows the peak is essentially a result of absorption in the bulk.

**Fig. 3** Photocurrent spectra obtained from the same GaAs wafer when the contact and light entrance surface is (a) the semi-insulating epilayer, (b) the substrate. (c) the spectrum in (a) (hollow circles) is reconstructed from the spectrum in (b) by multiplying it with the effect of absorption in the layer (solid curve).

Since the semiconductor essentially becomes opaque at the band-edge photon energy, and since this is a gradual process reaching its maximum effect at the band-edge, the observed photocurrent peaks will *always* appear *red-shifted* to the actual band-edge, i.e., the apparent peak position precedes the expected transition energy.

Intrinsic photon absorption continues to create electron-hole pairs, and excitons, even when the photon energy exceeds the gap (with the extra energy lost immediately to phonons). Since the same response continues at higher energies, the absorption spectrum should, in principle, follow a step. The same reasoning should also hold for spectra of other absorption-related features, such as intrinsic photoconductivity. On the other hand, exciton resonance peaks are sometimes observed in band-edge absorption spectra, mostly at low temperatures.[21] Consequently, when peaks have been observed in band-edge photocurrent spectra, the common practice has been to relate them to excitonic absorption. This seems to make sense because, like the exciton resonance peaks, they precede the bandgap and they are shaped as a peak. However, since excitons are electrically neutral, they cannot contribute directly to electric current. To be able to contribute, they need to dissociate. Dissociation mechanisms, such as defect assisted dissociation, impact ionization by free carriers, thermal dissociation, and Auger decay in neutral and charged impurities, have been invoked to support the idea of exciton-related photocurrent.[22,23] However, if this were indeed the case, then nanostructures, such as nanowires, wherein the mechanism we propose has no effect, should nonetheless show exciton-related peaks. Examining the literature of photocurrent measurements of, e.g., ZnO and GaN nanowires shows mostly step responses, as expected.[5,24,25] The exceptions that do show a peak are cases were the spectra have not been normalized to the photon flux and hence show a peak that *exceeds* the band-edge energy due to the diminishing spectrum of the lamp at the UV range.[26] We, therefore, sugesst, with all due caution, that if a photocurrent peak is observed in a bulk layer at room temperature, the effect of a bottom or a buried surface of the layer is to be suspected, before exciton resonances are invoked.

As evidenced above, the band edge response always precedes the actual bandgap due to the Frantz-Keldysh effect. Since this effect is a result of electric fields induced by charges trapped at surface states, the spectral data should contain information on the density of these surface charges. According to Eqs. 1 and 2, the measured current should be

$$I(h\nu) = I_D + (I_S - I_D)[1 - R(h\nu)]\exp\left(-\left(\frac{E_g - h\nu}{\Delta E}\right)^{3/2}\right) \quad (3)$$

where $I_D$ – dark current (or the current before the rise), $I_S$ – current after the rise, and $R(h\nu)$ – spectral reflectance from the surface of the sample. Rearranging Eq. 3, we get

$$y(h\nu) = \left[ln\left(\frac{I_S - I_D}{I(h\nu) - I_D}\right) + ln[1 - R(h\nu)]\right]^{2/3} = \frac{E_g - h\nu}{\Delta E} \quad (4)$$

The right-hand side of Eq. 4 is a linear expression. This means that if the suggested model is valid, then drawing *y(hν)* should yield a straight line intersecting the photon energy axis at the bandgap energy, providing a clear visual test to the validity of the model for the given data.

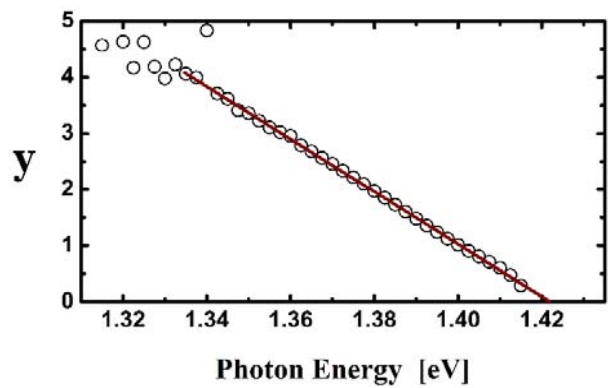

**Fig. 4** Applying a graphic method to obtain the bandgap from the data of Fig. 3b. Obtaining a straight line confirms the adequacy of the model for the specific data.

Figure 4 shows *y(hv)*, calculated from Fig. 3b using reflectance data from Phillip and Ehrenreich.[20] As expected, the data forms a straight line intersecting the photon energy axis at *1.4222±0.0097 eV*. Both the straight line and the bandgap value validate the adequacy of the model for this specific data. The slope of the line (along with the literature value of the reduced mass of GaAs) may be used to calculate the built-in electric field at the surface. We note, however, that the current model describes only cases where the data is the response of a single surface only, such as quantum wells or nanowires. Our GaAs wafer is an exception that falls in this category as well. Using Eq. 2, we get for the GaAs an electric field of *E=2.79·10$^5$ V/cm*. Using this result and the relation *εE=qN$_T$*, we get the *surface charge density* (or surface state density) *N$_T$=1.99·10$^{12}$ cm$^{-2}$*. This value is in agreement with previously reported value for native oxide covered GaAs(001) surface.[27] Similar analysis of the GaN nanowire of Fig. 2a yields *N$_T$=1.03·10$^{12}$ cm$^{-2}$* in agreement with values reported for the same nanowire device using a different method.[3] Dow and Redfield (1970) suggested that for excitonic transitions (as in GaN), the power of 3/2 in Eq. 1 reduces to 1.[28] Urbach tail is yet another mechanism that has been suggested to reduce the power to 1.[29] In such cases, if one still use a power of 3/2, the *y(hv)* should not follow a straight line. We did not observe such behavior, neither in InGaN/GaN quantum wells (not reported here), nor in GaN nanowires.

Cavallini *et al*. discusses several additional possible mechanisms other than field assisted absorption, such as structural disorder, defects, doping fluctuations, as well as broad excitonic, photonic, or plasmonic absorption, as alternative explanations for the red-shift.[3] These alternatives need to be excluded every time on a case by case basis. To this end, our proposed linear presentation of the data may serve as an easy-to-use method to validate that field assisted absorption is indeed the underlying mechanism in a given spectrum.

**V. Conclusion**

We presented a model accounting for the energy position variability of photocurrent band-edge features as resulting from variability in surface-state densities. We also presented a mechanism that may cause a peak-shaped response in thick layers and may not cause it in nanostructures. We suggested a graphical method to confirm the validity of the model for specific spectra and to accurately assess the transition energy (band-edge) and the surface-state charge density, adding quantitative features to photocurrent spectroscopy.